\makeatletter% added <<<<<<<<<<<<<<<<
\disable@package@load{program}{}
\makeatother
\documentclass[sn-mathphys,iicol]{sn-jnl}% Default with double column layout

\usepackage{braket}
\usepackage{tikz}
\usetikzlibrary{quantikz, arrows.meta, calc}
\usepackage{cleveref}
\usepackage{pgfplots}
\pgfplotsset{compat=newest}
\usepackage{stackengine}
\usepackage{mathdots}
\usepackage{multirow, multicol}
\usepackage{balance}

\newcommand{\hide}[1]{}

\newcommand{\N}{\mathbb{N}}
\newcommand{\C}{\mathbb{C}}
\newcommand{\R}{\mathbb{R}}
\newcommand{\aqc}{\textsc{aqc}}
\newcommand{\dwave}{\textsc{d-w}ave}
\newcommand{\qaoa}{\textsc{qaoa}}
\newcommand{\uqmc}{\textsc{uq}\maxcut}
\newcommand{\uqim}{\textsc{uqi}sing}
\newcommand{\spsa}{\textsc{spsa}}
\newcommand{\cobyla}{\textsc{cobyla}}
\newcommand{\adam}{\textsc{adam}}

\newcommand{\maxcut}{\textsc{m}ax\textsc{c}ut}

\newcommand{\mcal}[1]{\mathcal{#1}}
\newcommand{\mat}[1]{\mathbf{#1}}

\newcommand{\norm}[1]{\left| #1 \right|}
\newcommand{\Norm}[1]{\| #1 \|}
\newcommand{\Matrix}[2][0.8]{%
	\setlength{\arraycolsep}{5pt}\renewcommand{\arraystretch}{#1}
	\left[\hskip-5pt\begin{array}{*{16}{r}}#2\end{array}\hskip-5pt\right]}

 \newcommand{\graphinit}{
     \begin{tikzpicture}
        \foreach \a in {0,...,4}{
        \node[circle, fill=teal!50] (\a) at (\a*360/5+90: 3cm) {\a};
        }
        \foreach \a in {0,...,4}{
            \foreach \b in {\a,...,4}{
            \ifthenelse{\a=\b}{
                \ifthenelse{\a=0 \OR \a=4}{
                \node[above right] at (\a) {$C_{\a\b}$};
                }{
                \node[below left] at (\a) {$C_{\a\b}$};
                }
            }
            {
            \draw[-] (\a) edge node[fill=white]{$C_{\a\b}$} (\b);
            }
            }
        }
    \end{tikzpicture}
 }

 \newcommand{\graphopt}{
     \begin{tikzpicture}
        \foreach \a in {0,...,4}{
            \ifthenelse{\a=0 \OR \a=2}{
            \node[circle, fill=teal!50] (\a) at (\a*360/5+90: 3cm) {\a};
            }{
            \node[circle, fill=yellow] (\a) at (\a*360/5+90: 3cm) {\a};
            }
        }
        \foreach \a in {0,...,4}{
            \foreach \b in {\a,...,4}{
            \ifthenelse{\a=\b}{
                \ifthenelse{\a=0 \OR \a=4}{
                \node[above right] at (\a) {$C_{\a\b}$};
                }{
                \node[below left] at (\a) {$C_{\a\b}$};
                }
            }
            {
            \draw[-] (\a) edge node[fill=white]{$C_{\a\b}$} (\b);
            }
            }
        }
    \end{tikzpicture}
 }
\jyear{2021}%

\theoremstyle{thmstyleone}%
\theoremstyle{thmstyletwo}%

\theoremstyle{thmstyleone}%
\newtheorem{definition}{Definition}%

\begin{document}

\title[A Universal Quantum Algorithm for MaxCut and Ising Problems]{A Universal Quantum Algorithm 
for Weighted Maximum Cut and Ising Problems}


\author*[1]{\fnm{Natacha} \sur{Kuete Meli}}\email{meli@mic.uni-luebeck.de}

\author[1]{\fnm{Florian} \sur{Mannel}}\email{mannel@mic.uni-luebeck.de}

\author[1]{\fnm{Jan} \sur{Lellmann}}\email{lellmann@mic.uni-luebeck.de}

\affil*[1]{\orgdiv{Institute of Mathematics and Image Computing}, \orgname{University of L\"ubeck}, \orgaddress{\street{Maria-Goeppert-Straße 3}, \city{L\"ubeck}, \postcode{23562}, \state{Schleswig-Holstein}, \country{Germany}}}


\abstract{
%Finding the Weighted Maximum Cut 
%is a fundamental problem in combinatorial optimization and in statistical physics, where it relates to the Ising model used to describe ferromagnetism. 
We propose a hybrid quantum-classical algorithm to compute approximate solutions of binary combinatorial problems. 
We employ a shallow-depth quantum circuit to implement a unitary and Hermitian operator that block-encodes the weighted maximum cut or the Ising Hamiltonian. 
Measuring the expectation of this operator on a variational quantum state yields the variational energy of the quantum system. 
The system is enforced to evolve towards the ground state of the problem Hamiltonian by optimizing a set of angles using normalized gradient descent. 
Experimentally, our algorithm outperforms the state-of-the-art quantum approximate optimization algorithm on random fully connected graphs and challenges D-Wave quantum annealers by producing good approximate solutions. 
Source code and data files are publicly available under \href{https://github.com/nkuetemeli/UQMaxCutAndIsing}{https://github.com/nkuetemeli/UQMaxCutAndIsing}}.




\maketitle
\section{Introduction}
%%\paragraph{Quantum Computing.}
Quantum computing has emerged as a powerful computation paradigm taking advantage of principles of quantum mechanisms.
It involves two computational models that fundamentally differ in their functioning:
The adiabatic quantum model is best suited for optimization problems, typically cast in quadratic unconstrained binary optimization form. 
Commercial devices such as \dwave~annealers~\cite{Dwave} can already solve combinatorial problems in various fields such as computer vision \cite{benkner2021q,Meli2022} and database engineering \cite{groppe2021optimizing,uotila2022synergy}.
The universal model of quantum computing, also referred to as the gate-based or circuit model, is more flexible and can potentially implement any classical operation, as shown by Bennett \cite{bennett1973logical}.
For selected problems such as Shor's factoring \cite{shor1999polynomial} and Grover's search \cite{grover1996fast} algorithms, strong theoretical convergence properties and drastic speed-up of the universal quantum computing model over classical counterparts could be proven.
However, for problem sizes of practical interest, these algorithms still require more resources and run-time than existing universal devices can provide.
In the era of noisy intermediate-scale devices, it is a challenging task to find real-world applications of universal quantum computing.
Hybrid quantum-classical algorithms are therefore considered to be a promising way to obtain practical quantum supremacy.

%%\paragraph{Notation.}
%A qubit (quantum-bit) $\ket{\psi}$ is the fundamental unit of quantum information \cite{nielsen10,sutor2019dancing}. It is the state of a one-particle quantum system expressed as a superposition of a ground state $\ket{0}$ and an excited state $\ket{1}$ according to  $\ket{\psi} = \alpha \ket{0} + \beta \ket{1}$, with $|\alpha|^2 + |\beta|^2 = 1$, $\alpha, \beta \in \C$. The states $\ket{0} = (1, 0)^\top$ and $\ket{1} = (0, 1)^\top$ are the basis states of the system called eigenstates or stationary states. The qubit $\ket{\psi}$ irreversibly collapses to $\ket{0}$ or $\ket{1}$ with probability $|\alpha|^2$ or $|\beta|^2$ when it is measured. 
%We denote by $\ket{+}$ the uniform superposition state, i.e., $\ket{+} = (\ket{0} + \ket{1}) / \sqrt{2}$.
%%This definition of the state vector extends to an $n$-particle quantum system by considering the tensor product $\ket{\psi} = \otimes_{i=1}^n \ket{\psi_i} = \sum_{q=0}^{2^n-1} \alpha_q \ket{q}$ of the $n$ single-qubit-states. The eigenstates of the system are denoted by $\ket{q} \in \left\{ \ket{0}, \ket{1} \right\}^{\otimes n}$.

%\paragraph{The Problem.}
%This work aims to tackle the Ising spin model on universal quantum computers.

The Ising model \cite{mccoy2014two} describes a quantum mechanical system with $n\in \N$ particles or spins that can be in two possible states, i.e., the spin $s_i, i = 1, \ldots, n$ can be in the state $\pm 1$ 
%that we characterize by introducing the variable $q_i$ such that 
represented by $s_i = (-1)^{q_i}, q_i \in \left\{0, 1\right\}$.
Each spin $s_i$ can interact with some external energy of strength $\mcal{C}_{ii}$ or with an adjacent spin $s_j$ by a mutual interaction energy $\mcal{C}_{ij}$.
The complete system can be modelled by a general un-directed $ n $ vertices graph $ \mcal G = (\mcal S, \mcal E, \mcal{C}) $ with $ \mcal S = \left\{s_1, \ldots, s_n\right\} $, $ \mcal E \subseteq \mcal S \times \mcal S $ and a cost function $ \mcal{C} : \mcal E \to \R $ with $\mcal{C}(s_i, s_j) := \mcal{C}_{ij}$ on $ \mcal E $.
For a quantum system in the state $\ket{q} = \bigotimes_{i=1}^n\ket{q_i}, q_i \in \left\{0, 1\right\}$,
the Ising model describes the total energy of the system as being the expectation $\braket{\mat{C}} := \braket{q | \mat{C} | q}$ of the $(2^n \times 2^n)$ Hamiltonian
\begin{equation}
\label{eq:ising_model}
    \mat{C} = \sum_{i=1}^n \mcal{C}_{ii} \mat{Z}_i + \sum_{1 \leq i < j \leq n} \mcal{C}_{ij} \mat{Z}_i \mat{Z}_j,
\end{equation}
where $\mat{Z}_k$ denotes the Pauli-$\mat{Z}$ operator acting on the $k$th particle of the system.

The problem is to
\begin{equation}
\label{eq:initial_problem}
\begin{aligned}
& \underset{\ket{q}}{\text{minimize}}
& & \braket{\mat{C}} \\
& \text{subject to}
& & \ket{q} = \bigotimes_{i=1}^n\ket{q_i},  \; q_i \in \left\{0, 1\right\},
\end{aligned}
\end{equation}
i.e. to search for the state $\ket{q^\star}$ of the system with the minimal energy according to the Ising model in \cref{eq:ising_model}.
By setting $\mcal{C}_{ii} = 0$ for all $i$, the problem reduces to the so-called weighted maximum cut (\maxcut) problem \cite{goemans1995improved}.

The observable $\mat{C}$ in \cref{eq:ising_model} is a diagonal matrix and it is straightforward to verify that the expectation $\braket{\mat{C}}$ is lower-bounded by the smallest eigenvalue $\mcal{C}_{min}$ of $\mat{C}$. Indeed, measuring $\mat{C}$ on a quantum system prepared in the superposition state $\ket{\psi} = \sum_{q=0}^{2^n-1} \alpha_q \ket{q}$ yields
\begin{align}
    \braket{\mat{C}} = \braket{\psi | \mat{C} | \psi}
    &= \sum_{q=0}^{2^n-1} \norm{\alpha_q}^2 \braket{q | \mat{C} | q} \\
    &\geq \sum_{q=0}^{2^n-1} \norm{\alpha_q}^2 \braket{q^\star | \mat{C} | q^\star} = \mcal{C}_{min}.
\end{align}
In other words, under all norm-1 vectors, $\braket{\mat{C}}$ reaches the minimum exactly for $\ket{\psi}$ being an eigenvector of $\mat{C}$ to the lowest eigenvalue, a \emph{ground state} $\ket{q^\star}$ of $\mat{C}$.
Computing $\braket{\mat{C}}$ for all eigenvectors of $\mat{C}$---which amounts to computing the diagonal elements of $\mat{C}$---is practically hard, as it amounts to brute-forcing the problem. 
Instead, in hybrid quantum-classical approaches, one aims for an efficient way to search for parameters $\Theta \in \Gamma$ that solve the variational problem
\begin{equation}
\label{eq:variational_problem}
\begin{aligned}
& \underset{\Theta \in \Gamma}{\text{minimize}}
& & \mcal L (\Theta),
\quad
\mcal L (\Theta) := \braket{\psi (\Theta) | \mat{C} | \psi (\Theta)},
\end{aligned}
\end{equation}
where $\Gamma$ is a suitable parameterization of the search space.

The Ising spin model is a powerful tool describing ferromagnetism in statistical mechanics \cite{mccoy2014two}
as well as many practically relevant \textsc{np}-problems \cite{lucas2014ising}.
\maxcut~is no less important. It finds practical applications in varying fields such as computer vision \cite{kolmogorov2004energy}, data clustering \cite{ding2001min}, and communications network design \cite{barahona1988application}.
Thus, determining solutions of the Ising spin model or the \maxcut~problem, even approximately, is of great practical interest.

Adiabatic quantum computation (\aqc) relies on the adiabatic theorem \cite{Albash_2018,Jansen_2007} and solves the Ising problem by performing an adiabatic evolution of the quantum system from the known and easily-prepared ground state of an initial Hamiltonian to that of the problem Hamiltonian.
\dwave~\cite{Dwave} quantum annealers are intermediate-scale devices implementing \aqc.
It is an established fact that the run-time of \aqc~algorithms scales inversely with the energy gap between the ground state and the first excited state of the system Hamiltonian \cite{Albash_2018,Jansen_2007,zener1932non}.
This introduces the necessity of carefully designing the system Hamiltonian or to use spectral gap amplification techniques \cite{somma2013spectral}.
Another workaround is the conception of (hybrid) gate-based algorithms that use moderate quantum resources with outer-loop classical optimization \cite{farhi2014quantum}.

Quantum approximate optimization algorithms (\qaoa), firstly introduced by Fahri et al.~\cite{farhi2014quantum} and widely discussed in the literature \cite{guerreschi2019qaoa,hadfield2019quantum,herrman2021lower, shaydulin2019evaluating}, mimic the \aqc~model, but use a low-depth variational circuit capable to run on near-term devices and benefiting from the maturity of classical optimization.
\qaoa~is considered as one of the major candidates for dealing with realistic real-world applications at competitive performance using the universal model.
However, in practice, optimizing \qaoa~parameters appears to be extremely difficult due to a non-convex objective \cite{guerreschi2019qaoa,herrman2021lower, shaydulin2019evaluating}. 
Also, \qaoa~is three-folds iterative and hence computationally expensive: First, the ansatz itself involves repetitive layers of gates; second, the classical optimization routine used around \qaoa~often needs to be iterative as the objective function is non-convex; third, evaluating this objective function requires a large number of quantum circuit measurements.

%\paragraph{Contribution.}
This work aims to facilitate combinatorial optimization on universal quantum computers. 
We first eliminate the repetitive layers of \qaoa~and propose an alternative, more effective encoding of the problem on universal quantum hardware.
We present a new, easy to implement and low-depth variational circuit that block-encodes the \maxcut~or the Ising Hamiltonian in an Hermitian and unitary operator. 
Measuring this circuit reveals direct information about the loss function \cref{eq:variational_problem}.
We then derive an optimization routine based on gradient descent for the proposed variational circuit in order to drive the quantum system towards the ground state of the problem Hamiltonian.
We experimentally validate the novel algorithm and find that it outperforms the state-of-the-art gate-based \qaoa~model for solving \maxcut; on the Ising model, it also challenges the specialized \dwave~annealers.

This paper is structured as follows:
In \cref{sec:related_work} we recall how \dwave~and \qaoa~solve combinatorial problems;
in \cref{sec:presented_work}, we present the new variational quantum circuit and how its parameters are optimized;
\cref{sec:experiments} provides experimental results and \cref{sec:conclusion} concludes the work.

\section{Related Work}
\label{sec:related_work}
\subsection{Adiabatic Quantum Computing}
\label{sec:aqc}
Adiabatic quantum computation (\aqc) is an optimization paradigm which relies on the adiabatic theorem \cite{Jansen_2007, Albash_2018} saying that the ground state of a quantum mechanical system is the solution of an optimization problem \cite{transverse_field_qa_Kadowaki}.
Theoretically, the evolution of a quantum system of $n \in \N$ particles at a time $t \in [0, T]$, typically re-scaled to $s=t/T \in [0, 1]$, 
can be described by a Hamiltonian $\mat{H}(s)$ on a Hilbert space $\mcal H = \C^{\otimes 2^n}$, with the state of the system given by a unit vector $\ket{\psi(s)} \in \mcal H$. 

Industrial devices such as the \dwave~quantum annealer \cite{Dwave} have already been proposed to solve binary combinatorial problems based on the \aqc~optimization principle.
Their main idea consists in \emph{initializing} a quantum system with a Hamiltonian $\mat{B} = \sum_{k=1}^n \mat{X}_k$ whose ground state $\ket{+}^{\otimes n}$ is the perfect superposition state and easy to prepare. As above, $\mat{X}_k$ denotes the Pauli-$\mat{X}$ operator acting on the $k$th particle of the system. 
Then, another Hamiltonian $\mat{C}$, the \emph{problem Hamiltonian,} is prepared as in \cref{eq:ising_model}.
As the time $s$ evolves from $0$ to $1$, the initial Hamiltonian $\mat{B}$ is transformed into the problem Hamiltonian $\mat{C}$, describing a time-dependent system Hamiltonian
\begin{equation}
\label{eq:hamiltonian}
    \mat{H}(s) = \mcal{B}(s) \cdot \mat{B} + \mcal{C}(s) \cdot \mat{C},
\end{equation}
where $\mcal{B}$ and $\mcal{C}$ with $\lim_{s\to 1}\mcal{B}(s) = 0$ and $\lim_{s\to 1}\mcal{C}(s) = 1$ are annealing functions.
The evolution of the system generated by $\mat{H}(s)$ over the time $s \in [0, 1]$ is governed by the Schrödinger equation; its solution defines a time-dependent unitary operator that transforms the ground state of $\mat{B}$ into the ground state of $\mat{C}$ 
with high probability if $s$ varies sufficiently slowly \cite{Jansen_2007}. The ground state of $\mat{C}$ is the solution of the optimization problem \eqref{eq:initial_problem}.

While being an efficient scheme for combinatorial optimization that has the potential to ultimately supercede classical computers, \aqc~has a caveat. 
It is known that the smaller the energy gap between the ground state and the first excited state of the adiabatic Hamiltonian $\mat{H}(s)$, the longer the required annealing time for guaranteeing the success of the optimization \cite{Albash_2018,Jansen_2007,zener1932non}.
To overcome this, methods for universal quantum computers that take advantage of efficient classical optimization techniques have been proposed in the form of quantum approximate optimization algorithms \cite{farhi2014quantum}.

\subsection{Quantum Approximate Optimization Algorithms}
%Quantum approximate optimization algorithms (\qaoa), firstly introduced by Farhi et al.~\cite{farhi2014quantum} and widely extended and discussed in the literature \cite{hadfield2019quantum}, are variational algorithms mimicking adiabatic optimization for approximately solving combinatorial optimization problems in the form of \eqref{eq:initial_problem} on universal quantum computers.

In Quantum approximate optimization algorithms (\qaoa, \cite{farhi2014quantum}), the problem \eqref{eq:initial_problem} is \emph{relaxed} to finding a state $ \ket{\psi} = \sum_{q} \alpha_q \ket{q} $ such that $\braket{\mat{C}}$ is minimized.
\qaoa~solves the \maxcut~problem by \emph{trotterizing} the evolution generated by \cref{eq:hamiltonian}. Its pseudo-code is recapitulated in \cref{alg:qaoa}:
%and its circuit in \cref{fig:qaoa}:
First, the system is prepared in the perfect superposition state $\ket{+}^{\otimes n}$.
Then, trotterization consists in repeatedly, say $p$ times, applying %on the trial state $\ket{+}^{\otimes n}$ and
to the state the unitaries $ \mat{U}(\mat{C}, \gamma_k) :=  \exp(-i \gamma_k \mat{C}) $ and $ \mat{U}(\mat{B}, \beta_k) :=  \exp(-i \beta_k \mat{B}) $ generated by the problem Hamiltonian $\mat C$ and the mixing Hamiltonian $\mat B$. The (small) time steps $\gamma_k, \beta_k \in \R$, $1\leq k\leq p$, are optimization parameters in \qaoa.
The resulting state is called $ \ket{\gamma, \beta} $:
\begin{align}
\label{eq:qaoa}
    \ket{\gamma, \beta} &:=\nonumber \\
    \mat{U}(\mat{B}, &\beta_p)\mat{U}(\mat{C}, \gamma_p) \cdots \mat{U}(\mat{B}, \beta_1)\mat{U}(\mat{C}, \gamma_1)\ket{+}^{\otimes n}.
\end{align}
Finally, the state $ \ket{\gamma, \beta} $ is measured in the computational basis and is used to evaluate the cost function $\braket{\mat{C}} := \braket{\gamma, \beta|\mat{C}|\gamma, \beta}$ and if necessary its derivatives $\nabla_{(\gamma, \beta)}\braket{\mat{C}}$ and $\nabla_{(\gamma, \beta)}^2\braket{\mat{C}}$.
A classical optimization routine is used to update the parameters $ (\gamma, \beta) := (\gamma_1, \ldots, \gamma_p, \beta_1, \ldots, \beta_p) $.

\begin{algorithm}
\caption{Quantum Approximate Optimization Algorithms \cite{farhi2014quantum}}\label{alg:qaoa}
\begin{algorithmic}
\Require $ \mcal G = (\mcal S, \mcal E, \mcal C) $ and $p$
\Ensure $ \ket{\psi} = \ket{\gamma, \beta}$

\State Initialize the system in $ \ket{+} = \sqrt{2^{-n}} \sum_{q=0}^{2^n-1} \ket{q}$
\State Initialize parameter $ (\gamma, \beta) \gets (\gamma, \beta)^{init} $

\While{stopping criteria not met}
    \State Prepare $ \ket{\gamma, \beta}$ vis \cref{eq:qaoa}
    \State Measure $ \ket{\gamma, \beta} $ in the computational basis
    \State Compute $\braket{\mat{C}} := \braket{\gamma, \beta|\mat{C}|\gamma, \beta}$
    \State Update $(\gamma, \beta) \gets (\gamma, \beta)^{new}$ using a classical optimizer
\EndWhile
\end{algorithmic}
\end{algorithm}

%\begin{figure}[!ht]
%    \centering
%    \begin{tikzpicture}
%        \node[scale=.8]{
%        \begin{quantikz}
%        \lstick{$ \ket{0}_{q_1} $} & \qw & \gate{\mat{H}} & \gate[wires=4]{\mat{U}(\mat{C}, \gamma_1)} & \gate{R_x(\beta_1)} & \push{\ldots} & \gate[wires=4]{\mat{U}(\mat{C}, \gamma_p)} & \gate{R_x(\beta_p)} & \qw & \meter \qw & \cw\\
%        %
%        \lstick{$ \ket{0}_{q_2} $} & \qw & \gate{\mat{H}}  & & \gate{R_x(\beta_1)} & \push{\ldots} & \qw & \gate{R_x(\beta_p)} & \qw &\meter \qw & \cw\\
%        %
%        \lstick{$ \vdots \ $} &    &          & & & & & \\
%        %
%        \lstick{$ \ket{0}_{q_n} $} & \qw & \gate{\mat{H}} & & \gate{R_x(\beta_1)}  & \push{\ldots} & \qw & \gate{R_x(\beta_p)} & \qw & \meter \qw & \cw
%    \end{quantikz}
%    };
%    \end{tikzpicture}
%    \caption{A \qaoa~circuit used for the \maxcut~problem. First, the uniform superposition state $\ket{+}^{\otimes n}$ is prepared by applying a Hadamard gate to each qubit. Next, the state $\ket{\gamma, \beta}$ is prepared by transforming $\ket{+}^{\otimes n}$ by $p$ layers of parameterized unitaries $ \mat{U}(\mat{C}, \gamma_k) $ and $ \mat{U}(\mat{B}, \beta_k) $. Finally, $\ket{\gamma, \beta}$ is measured in the computational basis and used in a classical optimization routine to output the ground state $\ket{q^\star}$ of the \maxcut-Hamitonian with high probability.}
%    \label{fig:qaoa}
%\end{figure}

For $p \to \infty$ the results of \cite{farhi2014quantum} guarantee that there exist parameters $ (\gamma, \beta) $ for which measuring $ \ket{\gamma, \beta} $ gives the desired ground state $\ket{q^\star}$ with high probability.
However, the \qaoa~objective is difficult to optimize \cite{guerreschi2019qaoa,herrman2021lower, shaydulin2019evaluating}.
We believe that this is partially due to the fact that the \qaoa~ansatz encodes problem information in the argument of $\exp$ as phases of the qubits, which is partially lost at the measurement.
Another issue of \qaoa~is that its repetitive layers are still too expensive for running on current and near-term devices.
In this work, we propose a quantum circuit that encodes the problem more effectively and does not require the repetitive layers of \qaoa.
Adhering to the promising concept of designing hybrid quantum algorithms, we embed the circuit in a classical optimization method 
to output the desired ground state $\ket{q^\star}$ with high probability.

\section{Proposed Universal Quantum Algorithm}
\label{sec:presented_work}
Our method builds on the notion of \emph{block-encoding} introduced in \cite{gilyen2019quantum,Camps9951292} that
%consists in embedding properly scaled matrices that are not quantum logic gates as the principal block of a 
allows to embed non-unitary matrices as the principal block of a unitary operator acting on the quantum system. 
Block-encoding is typically achieved by enlarging the Hilbert space of the quantum system.

\begin{definition}[Block-encoding \cite{gilyen2019quantum,Camps9951292}]
  Let $a,n \in \N$ and $m:=a+n$. The $m\times m$ unitary $\mat{U}$ is said to be a block-encoding of the $n\times n$ matrix $\mat{C}$ if
  there is $\kappa\in (0,\infty)$ such that
  \begin{equation}
       \kappa \mat{C} = \left[\bra{0}^{\otimes a} \otimes \mat{I}_n\right] \mat{U} \left[\ket{0}^{\otimes a} \otimes \mat{I}_n\right].
  \end{equation}
\end{definition}

\noindent
By adding a single qubit to the system, our goal is to implement the $(2^{1+n}) \times (2^{1+n})$ \emph{unitary} operator $\mat{U} := \mat{U}(\mat{C}, K)$ given by
\begin{equation}
    \label{eq:U}
    \mat{U} :=
    \Matrix{
        \sin(\hat{\mat{C}}) & \cos(\hat{\mat{C}})\\
        \cos(\hat{\mat{C}}) & -\sin(\hat{\mat{C}})
    },\quad \hat{\mat{C}} := \mat{C} / K,
\end{equation}
for the Ising Hamiltonian $\mat{C}$ from \cref{eq:ising_model} and a suitably chosen \emph{constant} $ K \in \R$.
As $\mat{C}$ is a diagonal matrix, the $\sin$ and $\cos$ functions directly apply to the diagonal elements \cite[Section~2.1.8]{nielsen10}.
This allows to encode information about the problem as probability amplitudes of the qubits.
Note that because $\mat{C}$ is \emph{Hermitian}, $\mat{U}$ is Hermitian as well and thus can serve as measurement observable.
We use the constant $K$ to re-scale all entries of $\mat{C}$ to $[-\pi / 2, \pi / 2]$, where $\sin$ is strictly increasing and invertible. 
Specially, \cref{eq:U} block-encodes a bijective transformation of $\mat{C}$.
For reasons that will become clear in \cref{sec:circuit}, this re-scaling also allows for an efficient implementation of $\mat{U}$.
We stress that $K$ should not be set to large since $\lim_{K \to \infty} \mat{U}(\mat{C}, K) = \mat{X}^{\otimes(1+n)}$, i.e., for large $K$ the operator $\mat{U}$ behaves like a \textsc{not}-gate. 
In \cref{sec:optimization} we provide an optimization routine for the proposed circuit, 
we discuss its scalability and complexity in \cref{sec:scalability}, and lastly we provide a suitable choice for $K$ in \cref{sec:k_rescaling}.

\begin{figure*}[ht]
    \centering
    \begin{tikzpicture}
        \node[scale=.97]{
        \begin{quantikz}
        \lstick{{\sc Cost qubit} \\ {\footnotesize $\ket{\hat \psi}_c = \mat{X} \ket{0} $}} & & \lstick{$ \ket{\hspace{.05cm} \hat \psi \hspace{.05cm}}_c $} & \targ{} \gategroup[3,steps=5,style={dashed,teal!50,
        fill=teal!50, inner xsep=2pt},
background,label style={label position=above,anchor=
        north,yshift=.5cm}]{{\sc Quadratic term}} & \targ{} & \gate[style={fill=teal!50}]{R_y(-2\hat{\mcal{C}}_{ij})} & \targ{} & \targ{}  & \qw & \targ{} \gategroup[3,steps=3,style={dashed,teal!50,
        fill=teal!50, inner xsep=2pt},
background,label style={label position=above,anchor=
        north,yshift=.5cm}]{{\sc Unary term}} & \gate[style={fill=teal!50}]{R_y(-2\hat{\mcal{C}}_{ii})} & \targ{} & \qw\\
        \lstick[wires=2]{{\sc Working qubits}} & &\lstick{$ \ket{\hspace{.05cm} \cdot \hspace{.05cm}}_{q_i} $} & \qw &\ctrl{-1} & \qw &\ctrl{-1} & \qw  & \qw & \ctrl{-1} & \qw & \ctrl{-1} & \qw\\
        & & \lstick{$ \ket{\hspace{.05cm} \cdot \hspace{.05cm}}_{q_j} $} & \ctrl{-2} & \qw & \qw & \qw & \ctrl{-2} & \qw & \qw  & \qw & \qw & \qw
    \end{quantikz}
    };
    \end{tikzpicture}
    \caption{Implementation of the $(2^{1+n}) \times (2^{1+n})$ operator $ \mat{U} $. The cost qubit is initialized in the state $\ket{\hat \psi}_c = \mat{X}\ket{0}$ and is the target of all operations. Note that this does not contradict the idea of keeping it in the $\ket{0}$-state, as the operation $\mat{X} = R_y(\pi) \cdot \mat{Z}$ implements the last part of $\mat{U}$ (\cref{eq:U_quadratic}). 
    \textbf{(Left)} For each coupling edge of weight $\mcal{C}_{ij}$ between nodes $q_i$ and $q_j$, rotate the cost qubit by $R_y (-2\hat{\mcal{C}}_{ij})$ if their corresponding working qubits are in the same state, else by $R_y (2\hat{\mcal{C}}_{ij}) = \mat{X} \cdot R_y (-2\hat{\mcal{C}}_{ij}) \cdot \mat{X}$.
    \textbf{(Right)} For each unary edge of weight $\mcal{C}_{ii}$ involving node $q_i$, rotate the cost qubit by $R_y (-2\hat{\mcal{C}}_{ii})$ if its corresponding working qubit is in the $\ket{0}$-state, else by $R_y (2\hat{\mcal{C}}_{ii}) = \mat{X} \cdot R_y (-2\hat{\mcal{C}}_{ij}) \cdot \mat{X}$.}
    \label{fig:edge_circuit}
\end{figure*}
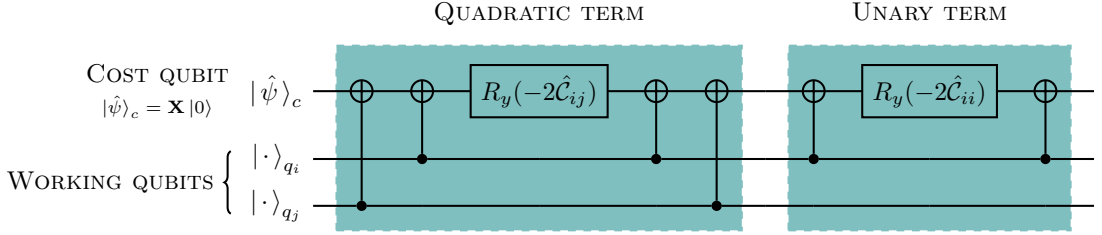

\subsection{Implementation}
\label{sec:circuit}
For a system prepared in the basis state $\ket{q}$, the cost of the cut $q$ according to the Ising model in \cref{eq:ising_model} is given by 
\begin{equation}
    \braket{\mat{C}} = \sum_{i=1}^n (-1)^{q_i} \mcal{C}_{ii}
    + \sum_{1 \leq i < j \leq n} (-1)^{q_i + q_j} \mcal{C}_{ij}.
\end{equation}
Crucially, for a $(1+n)$-qubit system prepared in the state $\ket{\hat \psi, \psi} := \ket{\hat \psi}_c \otimes \ket{\psi}$, where  $\ket{\hat \psi}_c = \alpha \ket{0}_c + \beta \ket{1}_c$ is a $1$-qubit register that we call the \emph{cost qubit} and $\ket{\psi}$ is the $n$-qubit \emph{working register,} 
%we crucially have for $\mat{U}$ defined in \cref{eq:U} that
%\begin{align}
%    \braket{\mat{U}} = 
%    \left( \norm{\alpha}^2 - \norm{\beta}^2 \right) \braket{\psi| \sin(\hat{\mat{C}}) |\psi} 
%    + \left(\alpha \bar \beta + \bar \alpha \beta\right) \braket{\psi| \cos(\hat{\mat{C}}) |\psi}.
%\end{align}
%Consequently, 
if the cost qubit $\ket{\hat \psi}_c$ is kept in the state $\ket{0}_c$, it holds for $\mat{U}$ defined in \cref{eq:U} that
\begin{equation}
    \label{eq:U_expectation}
    \braket{\mat{U}} = \braket{\psi|\sin\left[\dfrac{\mat{C}}{K}\right]|\psi}.
\end{equation}
Thus, if $K$ is chosen such that the entries of the scaled diagonal matrix $\frac{\mat{C}}{K}$ fit inside the monotone region of the sine, then minimizing $\braket{\mat{U}}$ with the cost qubit set to zero is equivalent to minimizing~$\braket{\mat{C}}$.

Observe that $\mat{U}$ is a block matrix of diagonal matrices and applying it to the basis state $\ket{0, q}$ 
%is equivalent 
gives the same result as applying to the cost qubit $\ket{0}_c$ the controlled $(2 \times 2)$-operator
\begin{equation}
    \mat{U}_{2\times 2} := 
    \Matrix{
        \sin\braket{\hat{\mat{C}}} & \cos\braket{\hat{\mat{C}}}\\
        \cos\braket{\hat{\mat{C}}} & -\sin\braket{\hat{\mat{C}}}
    }.
\end{equation}
As derived in \cref{app:U_geometric}, this operator performs a reflection in the Bloch sphere of the cost qubit about the axis 
$\vec n = (1/2) \Matrix{\cos\braket{\hat{\mat{C}}} & 0 & \sin\braket{\hat{\mat{C}}}}^\top$. 
Up to an irrelevant phase factor, the operator $\mat{U}$ can be written as
\begin{alignat}{3}
\label{eq:U_quadratic}
    \mat{U}(\mat{C}, K) 
     &\equiv R_y\left[2 \arccos (\sin\braket{\hat{\mat{C}}})\right] 
     &\cdot \mat{Z} \otimes \mat{I}^{\otimes n} \nonumber \\
     &\equiv R_y\left[\pi - 2 \braket{\hat{\mat{C}}}\right] &\cdot \mat{Z} \otimes \mat{I}^{\otimes n} \nonumber \\
     &\equiv R_y (-2\braket{\hat{\mat{C}}}) \cdot R_y(\pi) &\cdot \mat{Z} \otimes \mat{I}^{\otimes n}.
\end{alignat}
The second equation holds because $\arccos(\sin x) = \pi/2 - x$ for $x\in [-\pi/2, \pi/2]$.
The last equation is obtained by applying the identity $R_y(\theta_1 + \theta_2) = R_y(\theta_2)\cdot R_y(\theta_1)$ for rotations in two dimensions, where $\theta_1, \theta_2 \in \R$. 
By recursively applying this same identity to $R_y (-2\braket{\hat{\mat{C}}})$ we find
\begin{align}
\label{eq:U_quadratic_and_linear}
    \mat{U}&(\mat{C}, K) 
     \equiv \nonumber\\
     &\quad\prod_{i=1}^n \hspace{.8cm} \mat{X}^{q_i} \cdot R_y (-2\hat{\mcal{C}}_{ii}) \cdot \mat{X}^{q_i} \cdot \nonumber\\
     &\prod_{1 \leq i < j \leq n} \mat{X}^{q_i + q_j} \cdot R_y (-2\hat{\mcal{C}}_{ij})  \cdot \mat{X}^{q_i + q_j} \cdot \mat{X} \otimes \mat{I}^{\otimes n}.
\end{align}
As a result, the weighted sum of Pauli-$\mat{Z}$ operators naturally translates into a product of unitary transformations, which is very compatible with the gate-based model of quantum computing.
As the basis state $\ket{q}$ is chosen arbitrarily, $\braket{\mat{U}}$ outputs \emph{sine} transformed costs as derived in \cref{eq:U_expectation} for arbitrary states.
For a system prepared in the basis state $\ket{\hat{\psi}, \psi} = \ket{0, q}$, we can even recover the exact cost by $\braket{\mat{C}} = K \arcsin\braket{\mat{U}}$.

\noindent
The operator $\mat{U}$ can be efficiently implemented using the circuit given in \cref{fig:edge_circuit}:
\begin{itemize}
    \item We initialize the cost qubit in the state $\ket{\hat \psi}_c = R_y(\pi) \cdot \mat{Z}\ket{0} = \mat{X} \ket{0}$.
    \item For each weight $\mcal{C}_{ij}$ between two nodes $q_i$ and $q_j$, we rotate the cost qubit by $R_y (-2\hat{\mcal{C}}_{ij})$ if the corresponding working qubits are in the same state, and by $R_y (2\hat{\mcal{C}}_{ij}) = \mat{X} \cdot R_y (-2\hat{\mcal{C}}_{ij}) \cdot \mat{X}$ if they are not.
    \item For each unary weight $\mcal{C}_{ii}$ involving node $q_i$, we rotate the cost qubit by $R_y (-2\hat{\mcal{C}}_{ii})$ if the corresponding working qubit is in the $\ket{0}$-state, and by $R_y (2\hat{\mcal{C}}_{ii}) = \mat{X} \cdot R_y (-2\hat{\mcal{C}}_{ii}) \cdot \mat{X}$ if it is not.
\end{itemize}
Whenever the unary costs satisfy $\mcal{C}_{ii} = 0$ for all $i$, we refer to \cref{eq:U_quadratic_and_linear} as the Universal Quantum Maximum Cut (\uqmc) model, else as the Universal Quantum Ising (\uqim) model.

\subsection{Optimization}
\label{sec:optimization}

\subsubsection{Workflow}
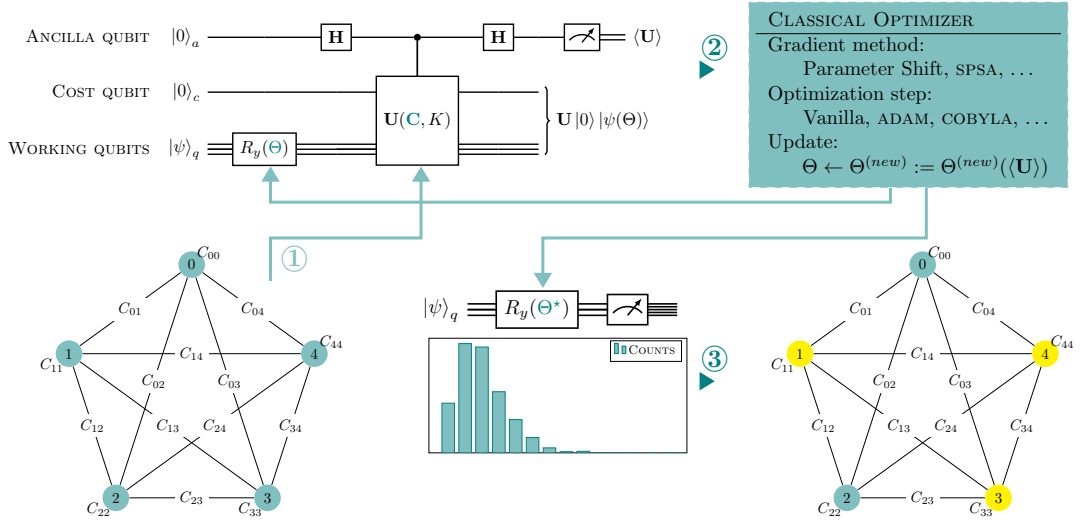
\begin{figure*}[]
    \centering
    
    \tikz \node[scale=.95]{
    
    \begin{tikzpicture}
    \node (complete_circuit) at (0, 0) [scale=.7]{
        \begin{quantikz}
        \lstick{{\sc Ancilla qubit}} & & \lstick{$ \ket{0}_a $}&\qw & \gate{\mat{H}} & \ctrl{+1} & \gate{\mat{H}} &\qw & \meter{} & \cw \rstick[wires=1]{$\braket{\mat{U}}$} \\
        \lstick{{\sc Cost qubit}} & & \lstick{$ \ket{0}_c $} &\qw  &\qw & \gate[wires=2][1.5cm]{\mat{U}(\textcolor{teal}{\mat{C}}, K)} \vqw{-1} & \qw  & \qw  \rstick[wires=5]{$\mat{U}\ket{0}\ket{\psi(\Theta)}$}  & \\
        \lstick[wires=1]{{\sc Working qubits}} & &\lstick{$ \ket{\psi}_{q} $} & \gate{R_y(\textcolor{teal}{\Theta})} \qwbundle[alternate]{} & \qwbundle[alternate]{} & \qwbundle[alternate]{} & \qwbundle[alternate]{} &\qwbundle[alternate]{}  &
    \end{quantikz}
    };
    \node (optimizer) at (8, 0) [scale=.8, dashed, draw=teal!50, fill=teal!50]{
        \begin{tabular}{l}
             {\sc Classical Optimizer} \\
             \hline
             Gradient method: \\
             \hspace{.5cm} Parameter Shift, \spsa , $\ldots$\\
             Optimization step: \\
             \hspace{.5cm} Vanilla, \adam, \cobyla, $\ldots$ \\
             Update: \\
             \hspace{.6cm}$\Theta \gets \Theta^{(new)} := \Theta^{(new)}(\braket{\mat{U}})$
        \end{tabular}
    };
    \node (graph_init) at (-2, -4) [scale=.6]{
        \graphinit
    };
    \node (measure_circuit) at (3.2, -3) [scale=.8]{
        \begin{quantikz}
        \lstick{$ \ket{\psi}_{q} $} & \gate{R_y(\textcolor{teal}{\Theta^\star})} \qwbundle[alternate]{} & \meter{}  \qwbundle[alternate]{} & \cw \qwbundle[alternate]{}  &
        \end{quantikz}
    };
    \node (counts) at (0.9, -5) [scale=.5]{
        \begin{axis}[
            ticks=none,
            tick align=outside,
            tick pos=left,
            xmin=-0.645, xmax=14.645,
            ymin=0, ymax=0.3,
            ybar, bar shift=0.5,
            % only scale the axis, not the axis including the ticks and labels
            scale only axis=true,
            % set `width' and `height' to the desired values
            width=0.45\textwidth,
            height=0.2\textwidth,
        ]
        
        \addplot[teal,fill=teal!50] table {counts.dat};
        \legend{\textsc{Counts}}
        
        \end{axis}
    };
    \node (graph_opt) at (8.2, -4) [scale=.6]{
        \graphopt
    };

    % Arrows
    \draw[-Triangle, very thick, teal!50](7.75, -1.3) -- (7.75, -1.5) -- (-.9, -1.5) -- (-.9, -1);
    \draw[-Triangle, very thick, teal!50](8.25, -1.3) -- (8.25, -2) -- (2.9, -2) -- (2.9, -2.7);
    
    \draw[-Triangle, very thick, teal!50] (-.9, -2.6) node[above right]{\raisebox{.5pt}{\textcircled{\raisebox{-.9pt} {$\textbf 1$}}}} 
    -- (-.9, -2) -- (1.2, -2) -- (1.2, -1.7)  -- (1.2, -1);
    \draw[-Triangle, very thick, teal] (5.2, .35) -- (5.3, .35) node[above]{\raisebox{.5pt}{\textcircled{\raisebox{-.9pt} {$\textbf 2$}}}};
    \draw[-Triangle, very thick, teal] (5.2, -4) -- (5.3, -4) node[above]{\raisebox{.5pt}{\textcircled{\raisebox{-.9pt} {$\textbf 3$}}}};

    \end{tikzpicture}
    };
    \caption{Complete workflow of our proposed algorithm for the Universal Quantum \maxcut~and Ising Model (\uqmc~and \uqim).
    \textbf{First} \raisebox{.5pt}{\textcircled{\raisebox{-.9pt} {$\mathbf{1}$}}}, the cost (node and edge weight) information for the given graph are used to implement the operator $\mat{U} := \mat{U}(\mat{C}, K)$ that block-encodes the problem Hamiltonians.
    This operator is applied to a trial variational quantum state $\ket{\psi(\Theta)}$ tensored with the cost qubit.
    \textbf{Second} \raisebox{.5pt}{\textcircled{\raisebox{-.9pt} {$\mathbf{2}$}}}, using the principle of implicit measurement, the expectation value $\braket{\mat{U}}$, which 
    equals the cost $\mcal{L}(\Theta)$, is approximated by measuring several copies of the circuit.
    This computed expectation and eventually its gradient are iteratively used in a classical optimization routine that drives the parameterized state $\ket{\psi(\Theta)}$ towards the state $\ket{\psi(\Theta^\star)}$ that potentially gives the global minimal cost value. 
    \textbf{Finally} \raisebox{.5pt}{\textcircled{\raisebox{-.9pt} {$\mathbf{3}$}}}, the optimal state $\ket{\psi(\Theta^\star)}$ is measured in the computational basis and the most frequently measured state corresponds to the desired optimal cut of the input graph.}
    \label{fig:complete_circuit}
\end{figure*}

The overall workflow of our algorithm is presented in \cref{fig:complete_circuit}.
The complete circuit consists of $3$ registers: a $1$-qubit register containing an ancilla qubit, another $1$-qubit register for the cost qubit, and an $n$-qubit register for the working qubits encoding the variables of the problem.

First, the working qubits are rotated by a set of angles $\Theta = (\theta_1, \ldots, \theta_n) \in \R^n$, constructing the ansatz
\begin{equation}
    \ket{\psi(\Theta)} := R_y(\theta_1) \otimes \cdots \otimes R_y(\theta_n) \ket{\psi},
\end{equation}
from a system previously prepared in the state $\ket{\psi}$. Here, the qubit $q_i$, representing the $i$th node of the graph, is rotated by $\theta_i$ around the $y$-axis.
Next, a Hadamard sandwich involving a controlled version of the $\mat{U}$-operator is applied to the ancilla qubit.
Finally, according to the principle of implicit measurement of quantum computing \cite[Section 4.4]{nielsen10} stating that \emph{all qubits that are not measured at the end of a quantum circuit can be assumed to be measured}, only the ancilla qubit is measured, leaving the cost and working qubits in the state $\mat{U} \ket{0, \psi(\theta)} = \ket{0}_c \otimes \sin (\hat{\mat C})\ket{\psi(\theta)} + \ket{1}_c \otimes \cos (\hat{\mat C})\ket{\psi(\theta)}$.

Our goal is to solve the variational problem
\begin{equation}
\label{eq:transformed_problem}
\begin{aligned}
& \underset{\Theta \in \R^n}{\text{minimize}}
& & \mcal L (\Theta),
\quad
\mcal L (\Theta) = \braket{0, \psi (\Theta) | \mat{U} | 0, \psi (\Theta)},
\end{aligned}
\end{equation}
i.e., to find a set of angles $\Theta^\star$ such that $\ket{\psi(\Theta^\star)} = \ket{q^\star}$.
The cost $\mcal L (\Theta)$ can be efficiently calculated by
$\mcal L (\Theta) = p(0) - p(1)$, where $p(0)$ and $p(1)$ are the probabilities of measuring the ancilla qubit in the $\ket{0}$ and $\ket{1}$ state, see \cref{app:implicit_measurement} for the proof. 
This allows us to compute the expectation $\braket{\mat{U}}$ without having to measure $\bra{0, \psi(\Theta)}$ needed for the scalar product.
For the optimization, several approaches \cite{mitarai2018quantum,spall1998overview} have fortunately been developed for evaluating or approximating gradients and improving optimization on quantum computers.

\subsubsection{Parameter Shift Rule}
The parameter shift rule \cite{mitarai2018quantum} is a simple but exact method for evaluating the analytical gradient of a function given in the form of an expectation as in \cref{eq:transformed_problem} on quantum hardware.
For computing the partial derivative with respect to the parameter $\theta_i$, it uses two function evaluations with the parameter $\theta_i$ shifted by $\pm \pi/2$:
\begin{equation}
    \dfrac{\partial}{\partial \theta_i} \mcal L (\Theta)
    = \dfrac{1}{2} \bigg[ \mcal L (\Theta_{i+}) - \mcal L (\Theta_{i-})\bigg],
\end{equation}
where $\Theta_{i\pm} := (\theta_1, \ldots, \theta_{i-1}, \theta_i \pm \pi/2, \theta_{i+1}, \ldots, \theta_n)$.
The parameter shift rule uses the same circuit as the function evaluation, but allows to compute the \emph{exact} gradient.
Its drawback is that it requires $2n$ function evaluations to compute the gradient of $\mcal L$. 
%the an $n$-dimensional problem as in our case. 

\subsubsection{Update Rule}
The circuits (\uqmc, \uqim) are optimized by normalized gradient descent with decreasing step size. 
Normalized gradient descent has recently been proposed by Suzuki et al.~\cite{suzuki2021normalized} as a powerful optimization strategy for variational quantum algorithms. 
Specifically, in \cite{suzuki2021normalized} it is demonstrated experimentally that normalized gradient steps are more effective in escaping non-global minima than gradient steps. It is also known \cite{MSS19} that normalized gradient descent evades saddle points. 
At each iteration $k$, our update rule reads
\begin{align}
\label{eq:update_rule}
    \Theta^{(k)} &\gets \Theta^{(k-1)} \nonumber\\
    &- \biggl[ \dfrac{\pi n}{2}\biggr]^{1/2}  \exp \Biggl[-\dfrac{4k^2}{k_{max}^2}\Biggr] 
    \cdot \dfrac{\nabla_\Theta \mcal L(\theta^{(k-1)})}{\Norm{\nabla_\Theta \mcal L(\theta^{(k-1)})}_2}.
\end{align}
The design of the update rule \cref{eq:update_rule} is motivated by the following consideration: 
We know that we produce bit-strings by either flipping $\ket{q_i}$ or not, thus $\theta_i^\star = \ell \pi, \ell \in \mathbb{Z}$.
At iteration $k=0$, the update rule \cref{eq:update_rule} allows each $\theta_i$ to get updated to $\theta_i^{1} \gets \theta_i^0 \pm \pi \cdot g_i/2$, where $ g_i \in [-1, 1]$ is the normalized contribution of $\theta_i^0$ in the loss $\mcal L (\Theta)$.
Subsequently, we let the step size decay exponentially to zero when approaching the maximum number of iterations $k_{max}$.
Given the noisy nature of quantum measurement, it is helpful that this frees us from the difficult task of determining a stopping condition for the algorithm.

\subsection{Scalability and Computational Complexity}
\label{sec:scalability}
\begin{table*}[]
    \centering
    \begin{tabular}{rcc}
    \hline
    \hline
        & \qaoa & \uqim~[Ours] \\
    \hline
      \# qubits                   & $\norm{\mcal S}$ & $\norm{\mcal S} + 2$ \\ 
      \# \textsc{cnot} gates      & $p\cdot(2\norm{\mcal{E}}  - 2|\mcal S|)$ &  $1+ 6|\mcal E| - 2|\mcal S|$ \\ 
      \# single qubits rotations  & $p\cdot(\norm{\mcal{E}} + \norm{\mcal S})$ & $2\norm{\mcal{E}}$ \\ 
      \# Hadamard gates          & $\norm{\mcal S}$ & $2$ \\ 
      qubit connectivity     & graph-dependent: $\norm{\mcal{E}} - \norm{\mcal S}$ & one-to-all: $\norm{\mcal S} + 1$ \\ 
      %classical cost evaluation 
      %                                & $\mcal{O}(\norm{\mcal{E}})$ operations  & $\mcal{O}(1)$ operations  \\ 
    \hline 
    \hline \\
    \end{tabular}
    \caption{Resources and complexity comparison of a depth $p$ conventional \qaoa~and our \uqim~model on a graph $ \mcal G = (\mcal S, \mcal E, \mcal{C}) $. 
    Recall that the set $\mcal{E}$ entails all the edges of the graph, i.e., the quadratic and unary ones. 
    The set $\mcal{S}$ is the set of vertices.
    Our construction requires less quantum and classical resources than \qaoa~for $p\geqslant 3$.}
    \label{tab:complexity}
\end{table*}

For a given graph $ \mcal G = (\mcal S, \mcal E, \mcal{C}) $, the circuit construction presented in \cref{fig:edge_circuit} requires at most one \textsc{not}-gate, $|\mcal E|$ single-qubit rotation gates and $4|\mcal E| - 2|\mcal S|$ \textsc{cnot} gates.
Note that since the edges can be treated in arbitrary order, two consecutive \textsc{cnot} gates that have the same control qubit cancel each other out as their product is the identity, further reducing the number of required \textsc{cnot} gates.

The controlled-$\mat{U}(\mat{C}, K)$ gate appearing in \cref{fig:complete_circuit} and conditioned by the ancilla qubit $\ket{\cdot}_a$ can be fully decomposed into single qubit rotations and \textsc{cnot} gates without using any Toffoli gates.
To see this, note that it can be expressed as %its matrix representation
\begin{align}
\label{eq:contolled_U_quadratic_and_linear}
    &\text{controlled-}\mat{U}(\mat{C}, K) 
    = 
    \mat{I} \otimes \mat{U}(\mat{C}, K)^{a} \nonumber\\
     &\equiv \mat{I} \otimes  \ 
     \prod_{i=1}^n \hspace{.8cm} \mat{X}^{q_i} \cdot [R_y (-2\hat{\mcal{C}}_{ii})]^a \cdot \mat{X}^{q_i} \cdot \nonumber\\
     & \quad  \quad \prod_{1 \leq i < j \leq n} \mat{X}^{q_i + q_j} \cdot [R_y (-2\hat{\mcal{C}}_{ij})]^a  \cdot \mat{X}^{q_i + q_j} \cdot \mat{X} \otimes \mat{I}^{\otimes n}.
\end{align}
In particular, in order to control the complete $\mat{U}(\mat{C}, K)$ gate, it suffices to control only the rotation gate $R_y$.
%, which 
%does not necessarily require the creation of Toffoli gates, but simply condition the rotation of the cost qubit by the ancilla qubit being in the state $\ket{1}$.

Finally, the controlled rotation itself can be decomposed into two \textsc{cnot} gates and two single qubit rotations as
\begin{equation}
    [R_y (\theta)]^a
    = \mat{X}^a \cdot R_y (-\theta/2) \cdot \mat{X}^a \cdot R_y (\theta/2).
\end{equation}

\Cref{tab:complexity} recapitulates the main differences between \uqim~and a conventional \qaoa~of depth $p$. 
It shows that for $p \geq 3$, \qaoa~requires more quantum resources than our \uqim~model. 
Further, physically mapping the \qaoa~ansatz onto the quantum hardware has to take into account a graph-dependent qubit connectivity, while our method, independently of the input graph, requires that one qubit (the cost qubit) is connected to all other qubits (ancilla and working qubits). 
%existing connections between the cost qubit and the remaining qubits.
%Another notable difference is the classical complexity for evaluating the cost function given measurement counts from the quantum computer: While \qaoa~requires storing a $2^n$-length real vector and performing $\mcal{O}(\norm{\mcal{E}})$ classical operations, our method measures only one qubit, stores $2$ real numbers, and evaluates the cost in $\mcal{O}(1)$ classical operations.

% ...requires storing counts for $2^n$ samples qubits

\subsection{Impact of the K-Rescaling}
\label{sec:k_rescaling}

When applying the method, it is important to appropriately choose the constant $K\in\R$ for re-scaling the costs $\hat{\mat{C}} = \mat{C}/K$.
The goal is to fix $K$ such that $\mathrm{diag}(\hat{\mat{C}}) \in [-\pi/2, \pi/2]^{2^n}$, as this
guarantees that $\sin(\mathrm{diag}(\hat{\mat{C}}))$ preserves the order of the original costs in $\mathrm{diag}(\mat{C})$.

\begin{figure}[ht]
    \centering
    \includegraphics[trim={.9cm, 0cm, 0cm, 0cm}, clip, scale=.42]{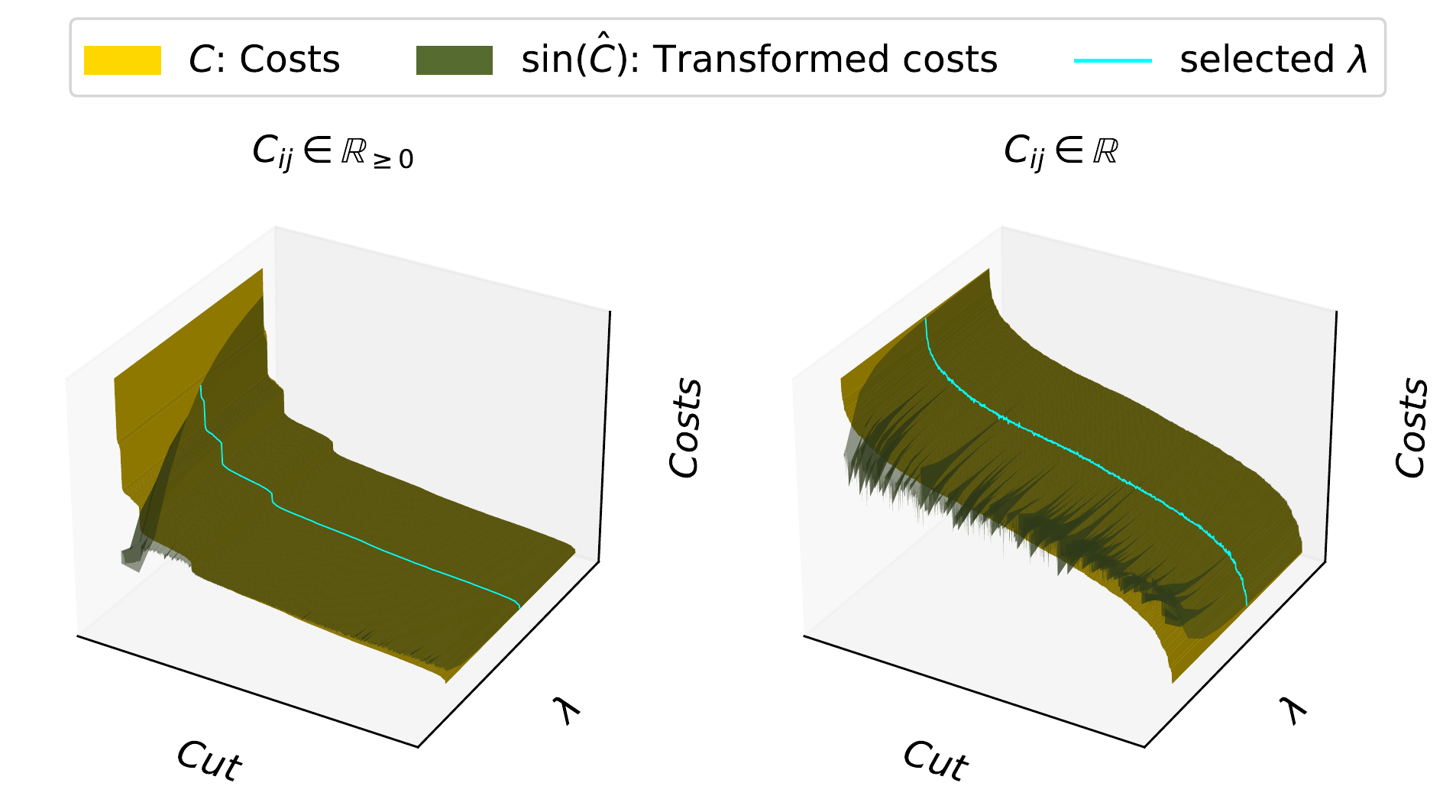}
    \caption{Transforming the true costs $\mathrm{diag}(\mat{C})$ into $\mathrm{diag}(\sin(\mat{C}/K))$, where $K = \lambda \cdot \mcal{C}_{max}$, c.f. \cref{eq:K}. 
    The true costs are sorted by their values and the $x$-axis represents the sorted cuts in the decimal basis. The $y$-axis showcases the influence of 
    $ 0.1 \leq \lambda \leq 1$. The $z$-axis reports the costs, re-scaled to $\left[0, 1\right]$. The transformed costs, sorted by the sorted arguments of the true costs, show whether the order is preserved under the influence of $\lambda$. Example: If $\mat{C} = [2, 5, 7, 3]$ and, for some value of $K$, $\mat{\hat{C}} = [1, 3, 4, 2]$, then $\mathrm{sort}(\mat C) = [2, 3, 5, 7]$, $\mathrm{argsort}(\mat C) = [0, 3, 1, 2]$ and $\mat{\hat{C}}_{\mathrm{sorted}} = [1, 2, 3, 4]$. 
    For each value of $\lambda$, the displayed costs are the averages over $10$ instances of a $10$-nodes fully-connected graph. 
    \textbf{Left:} Graph instances with only positive edge weights. \textbf{Right:} Graph instances with both positive and negative edge weights.
    As expected, the more $\lambda$ and thus $K$ grows, the better the order of the true cost agrees with the order of the transformed costs. Also, for fixed $\lambda$ the difference between the two costs is the largest when the true cost is large in absolute value. 
    The cyan-marked line is the profile that we selected for the experiments; it indicates the value $\lambda = 2/\pi$, which is the minimum $\lambda$ that guarantees an error-free transformation, cf. \cref{eq:K}.}
    \label{fig:experiment_K}
\end{figure}

Although it is tempting to choose $K \gg \mcal{C}_{max} := \max_k \mat{C}_{kk}$, where $\mat{C}_{kk}$ denotes the $k$th diagonal element, recall that in \cref{eq:U} we have $\lim_{K \to \infty} \mat{U}(\mat{C}, K) = \mat{X}^{\otimes(n+1)}$, i.e., the larger $K$, the more shots are required to accurately measure the entries of $\sin(\hat{\mat{C}})$. 
Fortunately, in many problems an upper bound to the maximal cost $\mcal{C}_{max}$ can be computed from the original weights without knowing $\mat{C}$. For example, choosing
\begin{equation}
\label{eq:K}
    K = \dfrac{2}{\pi} \cdot \mcal{C}_{max} := \dfrac{2}{\pi} \cdot \left[ \sum_{i=1}^n \norm{\mcal{C}_{ii}} + \sum_{1 \leq i < j \leq n} \norm{\mcal{C}_{ij}} \right]
\end{equation}
guarantees an error-free transformation of the initial problem \eqref{eq:variational_problem} into the equivalent problem \eqref{eq:transformed_problem}.
As presented for $10$-node graph examples in \cref{fig:experiment_K}, the choice of smaller values for $K$ entails that the approximation of $\arccos(\sin x)$ by $\pi/2 - x$, used in \cref{eq:U_quadratic}, is inaccurate, with more pronounced error in the largest absolute entries of $\mathrm{diag}(\hat{\mat{C}})$.
Notably, the solution of the transformed problem will only be a solution of the initial problem as long as $\arg \min_k \hat{\mat{C}}_{kk} = \arg \min_k \mat{C}_{kk}$.

\section{Experimental Results}
\label{sec:experiments}
In order to validate the practical usefulness of \uqmc~and \uqim, we benchmark against two state-of-the-art approaches for solving binary combinatorial optimization with quantum computing: \qaoa~for the gate-based model and \dwave~solvers for the adiabatic model.
Random graphs in the experiments are generated using the \textsc{Python} language package \textsc{NetworkX} \cite{hagberg2008exploring}.
The unary and quadratic edge weights are all randomly and uniformly chosen in the range $[1, 10]$ and the graphs are all fully connected.
The gate-based circuits in the experiments (\uqmc, \uqim, \qaoa) are implemented in \textsc{Python} and simulated in a noise-free framework using the \textsc{QisKit} library and the IBM-QASM simulator \cite{cross2018ibm}.
For the adiabatic model, \dwave~solvers that run on the actual quantum hardware are used.
\dwave~quantum annealers are made available through the Leap quantum cloud service \cite{Dwave_leap}, and the \dwave~quantum algorithms can be implemented in Python using the Ocean software \cite{Dwave_ocean}.
We perform $1024$ measurement shots for all gate-based algorithms.
On \dwave~\cite{Dwave}, the experiment is run with the default annealing time of $20\mu s$ and $50$ sample reads on the Advantage topology.

\subsection{Benchmark Metrics}
We denote by $\ket{q^\star}$ the ground-truth global minimizer and by $\ket{\psi^\star}$ the ground state proposal of each method (\uqmc/\uqim, \qaoa~and \dwave).
In the experiments, we adopt the following two metrics to evaluate the performance of the methods:
\begin{itemize}
    \item The approximation ratio 
    \begin{equation}
        r(\psi^\star) := \dfrac{\braket{\psi^\star | \mat{C} | \psi^\star} - \mcal{C}_{max}}{\mcal{C}_{min}-  \mcal{C}_{max}}
    \end{equation}
informs about the quality of the result, i.e., how confident the method is with its solution proposal and how far the cost of this proposal is from the minimal cost $\mcal{C}_{min} := \min_k \mat{C}_{kk}$, cf.~\cite{willsch2020benchmarking}. All the terms appearing in $r$ are classically evaluated. It holds $0 \leq r(\psi^\star) \leq 1$ and $r(\psi^\star)=1$ iff $\psi^\star = q^\star$.

    \item The approximation index
    \begin{equation}
        i(\psi^\star) := \mathbf{1}_{\braket{q^\star | \mat{C} | q^\star} = \braket{q_{max} | \mat{C} | q_{max}}
        %q^\star = q_{max}
        }
    \end{equation}
is a Boolean variable that indicates whether the most likely state $\ket{q_{max}}$ obtained with probability $\norm{\alpha_{max}}^2$ is the desired state $\ket{q^\star}$. 
It holds $i(\psi^\star) = 1$ if $\braket{q^\star | \mat{C} | q^\star} = \braket{q_{max} | \mat{C} | q_{max}}$ and $0$ otherwise.
Note that this differs from the usual approach of \dwave, where the sampled state with the minimal energy is regarded as the best solution proposal.
\end{itemize}

\subsection{\mbox{Benchmarking UQMaxCut}\-~\mbox{against QAOA and D-Wave}}
\subsubsection{Symmetric Solutions and Entanglement}
\begin{figure}[!h]
    \centering
    \begin{tikzpicture}
        \node[scale=.7]{
        \begin{quantikz}
        \lstick{$ \ket{0}_{q_1} $} & \ctrl{4} & \push{\cdots} & \ctrl{2} & \ctrl{1} & \gate{\mat{H}} & \ctrl{1} & \ctrl{2}& \push{\cdots} & \ctrl{4} & \qw\\
        \lstick{$ \ket{0}_{q_2} $} & \qw & \push{\cdots} & \qw & \targ{} & \qw & \targ{} & \qw  & \push{\cdots} & \qw  & \qw\\
        \lstick{$ \ket{0}_{q_3} $} & \qw & \push{\cdots} & \targ{} & \qw & \qw & \qw & \targ{} & \push{\cdots} & \qw  & \qw\\
        \lstick{$ \vdots \ $} & & & & & & & & & & &\\
        \lstick{$ \ket{0}_{q_n} $} & \targ{} & \push{\cdots} & \qw & \qw & \qw & \qw & \qw & \push{\cdots} & \targ{} & \qw
    \end{quantikz}
    };
    \end{tikzpicture}
    \caption{Entanglement circuit to enforce symmetric solutions for \maxcut. 
    Using this circuit, the optimization is performed only on the last $n-1$ variables and the qubit for the node $q_1$ is kept constant in the state $\ket{0}$ or $\ket{1}$.}
    \label{fig:entanglement}
\end{figure}
Before discussing the results on the \maxcut~problem, it is important to notice that
for \maxcut, solutions always exist in symmetric pairs. Specifically, for a basis state solution $\ket{q^\star} = \bigotimes_{i=1}^n\ket{q_i^\star}, q_i^\star \in \left\{0, 1\right\}$, the basis state $\ket{\bar{q}^\star} := \bigotimes_{i=1}^n\ket{1-q_i^\star}$ is also a solution. 
This feature can be enforced by introducing the entanglement circuit given in \cref{fig:entanglement} after the rotation layer in \cref{fig:complete_circuit}. 
The circuit has the matrix representation 
\begin{equation}
    E = \dfrac{1}{\sqrt{2}}\Matrix{
         1&      & & &       &1  \\
          &\ddots& & &\iddots&   \\
          &      &1&1&       &   \\
          &      &1&-1&      &   \\
          &\iddots& & &\ddots&   \\
         1&      & & &       &-1 \\
    }.
\end{equation}
Also, this entanglement allows to optimize over $n-1$ angles instead of $n$.
Without entanglement our method just outputs one of the two possible solutions.
In the experiments, \uqmc~is used without entanglement; if the algorithm outputs either one of the two solutions, we set the approximation index to~$1$.

\subsubsection{Benchmark Results}
\begin{figure*}[ht]
    \centering
    \includegraphics[scale=.45]{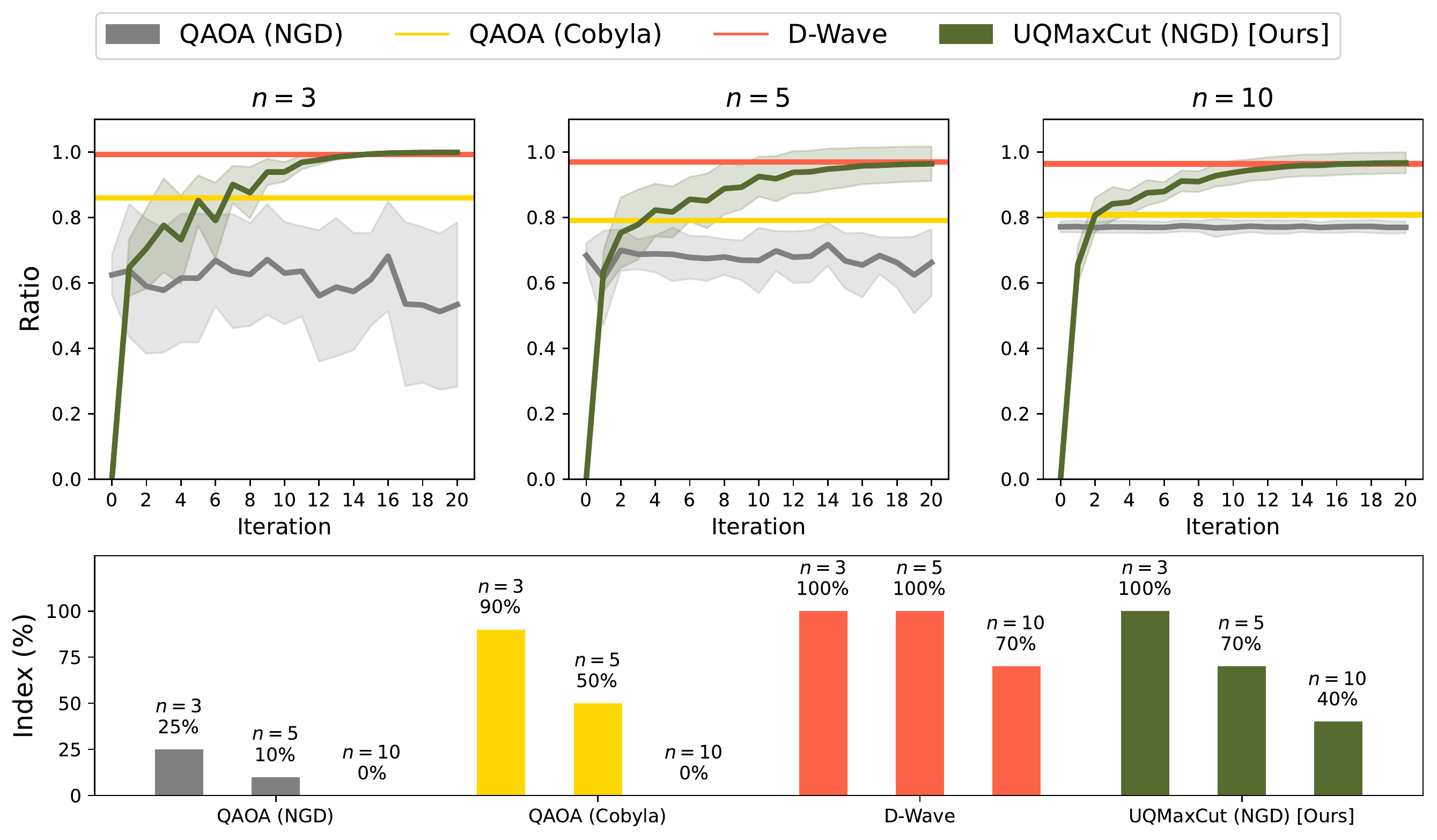}
    \caption{Experimental comparison of the proposed Universal Quantum \maxcut~(\uqmc) algorithm with \qaoa~and~\dwave.
    Results are shown for fully connected graphs of $n = 3, 5$ and $10$ nodes with strictly positive edge weights. They are averaged over $20$ random graph instances for each $n$.
    \textbf{Top:} The approximation ratio. The lines represent the averaged ratios over the $20$ instances and the shaded areas indicate the standard deviations.
    \textbf{Bottom:} The approximation index, i.e., the percentage of instances where a global solution is found.
    All gate-based methods hardly return a global solution for larger $n$. Yet, in contrast to \qaoa~and competitively with \dwave, the proposed \uqmc~consistently returns very good approximate solutions, i.e., points whose function values are very close to the global minimal function value.}
    \label{fig:experiment_benchmark_mc}
\end{figure*}
For the outer optimization algorithm of the \uqmc~circuit we use normalized gradient descent as described in \cref{sec:optimization}. 
Other optimization algorithms such as vanilla gradient descent or adaptive moment estimation (\textsc{adam} \cite{kingma2014adam}) could be used as well.
However, it proved difficult in the experiments to find suitable step sizes for those methods, so we leave them for future research.

The \qaoa~layers depth in the experiments is set to $p = \lceil n/2 \rceil $ 
to allow for a fair comparison, as then both \qaoa~and \uqmc~optimize over approximately $n$
real variables.
We also attempted to optimize \qaoa~using the same optimizer as for \uqmc, but the results were not competitive.
Hence, we also show the results of \qaoa~when using a gradient-free optimizer; we used the \textsc{cobyla} solver \cite{powell1994direct} available in the \textsc{scipy} library \cite{virtanen2020scipy}.

The results are presented in \cref{fig:experiment_benchmark_mc} for fully connected graphs of $n=3, 5$ and $10$ nodes with strictly positive edge weights. 
For each $n$, the results are averaged over $20$ graph instances and all algorithms are tested on the same instances.
The angles for \uqmc~and \qaoa~are all initialized to $0$.

The approximation ratio for the three methods (\qaoa, \dwave, \uqmc) is not adversely affected by the number of variables $n$, but the approximation index
%-- i.e., the fraction of runs where the correct global solution is found -- 
drops sharply as the size of the problem increases.
The gradient-free Cobyla-optimized version of \qaoa~performs much better than \qaoa~with gradient descent.
The latter is on average as good as \dwave~regarding the approximation ratio.
We conjecture that the gradient-based optimization of \qaoa~often gets trapped by saddle points of the \qaoa~loss function landscape.
In contrast, \uqmc~clearly outperforms the two \qaoa~variants and can compete with \dwave~in producing good approximate solutions. 
Furthermore, the approximation index demonstrates that it returns a global minimizer significantly more often than \qaoa~and less often than \dwave~whose architecture is specifically designed to solve such problems.

\subsection{Benchmarking UQIsing~against D-Wave}

\begin{figure*}[ht]
    \centering
    \includegraphics[scale=.45]{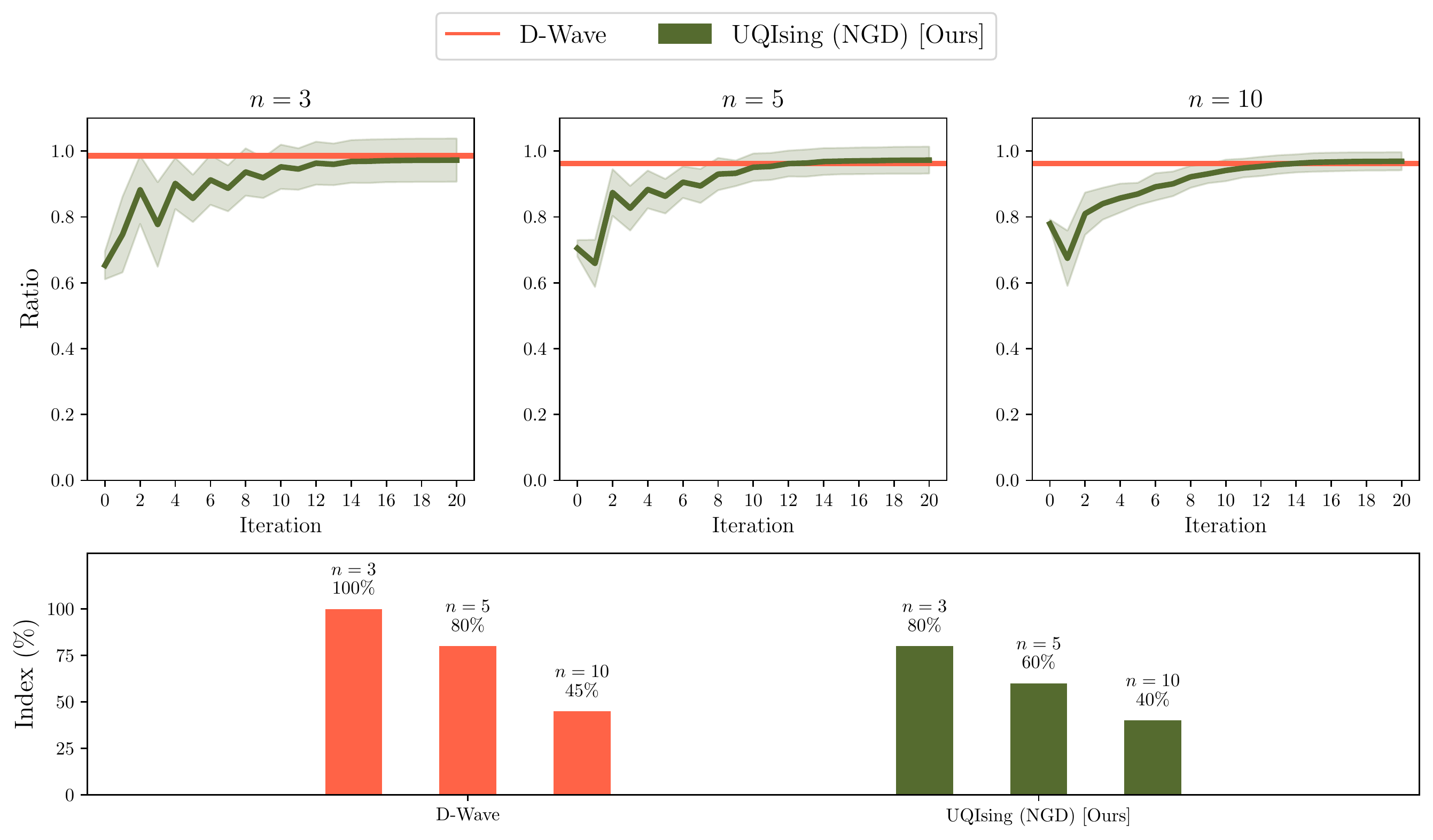}
    \caption{Experimental comparison of the proposed Universal Quantum Ising Model (\uqim) algorithm with ~\dwave.
    Results are shown for fully connected graphs of $n = 3, 5$ and $10$ nodes with strictly positive edge weights. 
    They are averaged over $20$ random graph instances for each $n$.
    \textbf{Top:} The approximation ratio. The lines represent the averaged ratios over the $20$ instances and the shaded areas indicate the standard deviations.
    \textbf{Bottom:} The approximation index, i.e., the percentage of instances where a global solution is found. 
    Our method can compete with \dwave~solvers in predicting approximate solutions and finding the global minimum for moderate $n$.}
    \label{fig:experiment_benchmark_ising}
\end{figure*}

For the Ising model we benchmark the proposed \uqim~algorithm against the \dwave~annealer, the adiabatic quantum computer specialized in solving this type of problems. 
The variational circuit for \uqim~is optimized in the same way as for \uqmc, see \cref{sec:optimization}, with the exception that the initial angles are set to $\pi/2$ instead of $0$.
The results are depicted in \cref{fig:experiment_benchmark_ising}.

Notable differences in performance between \uqim~and \dwave~are consistent with the MaxCut experiments.
Specifically, its high approximation ratio, similar to that of \dwave, indicates that \uqim~always produces either globally optimal solutions or extremely good approximations thereof. 
%In contrast, the small approximation ratio of \dwave~reveals that if the annealer does not sample the correct solution, then it may output a state with a comparably high energy value.
On the other hand, the approximation index shows that \dwave~identifies a globally optimal solution more often than \uqim.

\section{Conclusion}
\label{sec:conclusion}
We have presented a new low-depth quantum circuit to tackle two important combinatorial problems on universal quantum machines. 
The resulting Universal Quantum \maxcut~(\uqmc) approach outperforms the state-of-the-art quantum approximate optimization algorithms (\qaoa) by the lower depth, by the computed approximation ratios and by a higher probability of outputting optimal solutions. 
It also challenges the \dwave-quantum annealers that are specifically designed to solve such combinatorial problems; 
on the \maxcut~as well as on the Ising spin model, \uqmc, respectively, \uqim~achieve better approximation ratios and can compete with \dwave~in producing globally optimal solutions.

We believe that the proposed approach enables the design of new methods for solving practically-sized problems on universal quantum machines.
Inspired by the novel operator $\mat{U}$, future work should focus on designing fully universal algorithms without the classical outer optimization loop, replacing the latter with fully universal methods like for example Grover's search \cite{grover1996fast}.

\balance
\bibliography{sn-bibliography}% common bib file


\newpage
\begin{appendices}
\begin{figure*}
\section{Geometric Interpretation of \texorpdfstring{\(\mat{U}_{2\times 2}\)}{U}}
\label{app:U_geometric}
We want to show that the operator
\begin{equation}
    \mat{U}_{2\times 2} = 
    \Matrix{
        \sin\braket{\hat{\mat{C}}} & \cos\braket{\hat{\mat{C}}}\\
        \cos\braket{\hat{\mat{C}}} & -\sin\braket{\hat{\mat{C}}}
    }
\end{equation}
is up to a phase factor a reflection in the Bloch sphere about the axis 
$\vec n = (1/2) \Matrix{\cos\braket{\hat{\mat{C}}} & 0 & \sin\braket{\hat{\mat{C}}}}^\top$. More precisely, we show that 
\begin{equation}\label{eq_Uisreflection}
	\mat{U}_{2\times 2} \equiv R_y\Bigl[2 \arccos (\sin\braket{\hat{\mat{C}}})\Bigr]\cdot \mat{Z}.
\end{equation}

\textbf{Proof:}
For $\sin\braket{\hat{\mat{C}}} = \pm 1$, we have $\mat{U}_{2\times 2} = \pm\mat{Z}$, so
\cref{eq_Uisreflection} holds. In the remainder of the proof we can therefore assume
$\sin\braket{\hat{\mat{C}}} \neq \pm 1$. 
The matrix $\mat{U}_{2\times 2}$ has the spectral decomposition $\mat{U}_{2\times 2} = \mat{X}^\top \mat{\Lambda} \mat{X}$, where $\mat{\Lambda} = \text{diag}([-1 \; 1])$ is the diagonal matrix of eigenvalues of $\mat{U}_{2\times 2}$ and $\mat{X} = \Matrix{\ket{\psi_{\_}}  \; | \; \ket{\psi_{+}}}$ is the matrix containing its eigenvectors
\begin{align}
	\ket{\psi_{+}} &= \dfrac{1}{\sqrt{2-2\sin\braket{\hat{\mat{C}}}}}
	\Matrix{-\cos\braket{\hat{\mat{C}}} \\ \sin\braket{\hat{\mat{C}}} - 1} \\
	\qquad \text{and} \qquad
	\ket{\psi_{\_}} &= \dfrac{1}{\sqrt{2 + 2\sin\braket{\hat{\mat{C}}}}}
	\Matrix{-\cos\braket{\hat{\mat{C}}} \\ \sin\braket{\hat{\mat{C}}} + 1}.
\end{align}
Let $\mat{\Sigma} = \text{diag}([\pi \; 2\pi])$, so $\mat{\Lambda} = \exp{(i\mat{\Sigma})}$. The generalization of the exponential map to normal matrices \cite[Section~2.1.8]{nielsen10} implies that $ \mat{U}_{2\times 2} = \exp(i \mat{X}^\top \mat{\Sigma} \mat{X}) $. 
Now we express $\mat{X}^\top \mat{\Sigma} \mat{X}$ in the basis formed by the Pauli matrices:
\begin{align}
	\mat{X}^\top \mat{\Sigma} \mat{X} &= \dfrac{1}{2} 
	\Matrix{3\pi + \pi \sin\braket{\hat{\mat{C}}} 
		& \pi \cos\braket{\hat{\mat{C}}} \\
		\pi \cos\braket{\hat{\mat{C}}}
		& 3\pi - \pi \sin\braket{\hat{\mat{C}}}} \\
	&= \dfrac{3\pi}{2} \cdot \mat{I} + \dfrac{\pi}{2} \cos\braket{\hat{\mat{C}}} \cdot \mat{X} + 0 \cdot \mat{Y} + \dfrac{\pi}{2}\sin\braket{\hat{\mat{C}}} \cdot\mat{Z}.
\end{align}
We can thus rewrite $\mat{U}_{2\times 2}$ as $\mat{U}_{2\times 2} = \exp(i \alpha ) \exp(i \vec v \cdot \vec{\mat{\Sigma}})$, where $\alpha = 3\pi / 2$ and \\ $\vec v = (1/2) 
	\Matrix{\pi \cos\braket{\hat{\mat{C}}} & 0 & \pi \sin\braket{\hat{\mat{C}}} }^\top$.
The notation $\vec v \cdot \vec{\mat{\Sigma}} := v_x \cdot \mat{X} + v_y \cdot \mat{Y} + v_z \cdot \mat{Z}$ denotes a linear combination of Pauli matrices.
With a slight change of variables $\theta = -2 \Norm{\vec{v}}_2 = - \pi$ and $\vec n 
= \vec v / \Norm{\vec{v}}_2 =
\Matrix{\cos\braket{\hat{\mat{C}}} & 0 & \sin\braket{\hat{\mat{C}}}}^\top$, 
this is equivalent to
$\mat{U}_{2\times 2} = \exp(i \alpha ) \exp(-i (\theta/2) \vec n \cdot \vec{\mat{\Sigma}})$.

The global phase $\exp(i \alpha )$ can be ignored in the implementation as it has no effect on the measurement. The exponential term $\exp(-i (\theta/2) \vec n \cdot \vec{\mat{\Sigma}})$ corresponds to a rotation of angle $\theta$ of the Bloch sphere about the axis $\vec{n}$. As $\vec{n}$ lies in the $xz$-plane and we have $\theta = -2 \Norm{\vec{v}}_2 = -\pi$, it follows that $\mat{U}_{2\times 2}$ is a reflection about $\vec{n}$. Equivalently,
\cref{eq_Uisreflection} holds.
\vspace{4cm}
\end{figure*}

\newpage
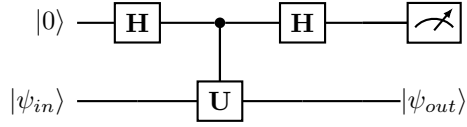
\begin{figure*}
\section{Implicit Measurement}
\label{app:implicit_measurement}
Let $\mat{U}$ be a Hermitian and unitary operator, so $\mat{U}$ can be used both as quantum gate and measurement observable. 
According to the implicit measurement principle of quantum computation~\cite[Section 4.4]{nielsen10}, \emph{all qubits that are not measured at the end of a quantum circuit can be assumed to be measured}. 
Specially, the measurement of the first qubit of the circuit in \cref{fig:implicit_measurement} performs a measurement of the observable $\mat{U}$ for a system prepared in the state $ \ket{\psi_{in}} $.

%\begin{figure}[h]
{
    \centering
    \begin{quantikz}
    \lstick{$ \ket{0} $} & \gate{\mat{H}} & \ctrl{+1} & \gate{\mat{H}} & \qw & \meter \qw &   \\
    \lstick{$ \ket{\psi_{in}} $} & \qw & \gate{\mat{U}} \vqw{-1} & \qw & \qw &   \push{ \ket{\psi_{out}} }  & 
    \end{quantikz}
    \caption{Circuit illustrating the implicit measurement principle.}
    \label{fig:implicit_measurement}
}
%\end{figure}

\textbf{Proof:} %One can easily check that $\mat{U}$ has the
As $\mat{U}$ is both Hermitian and unitary, its eigenvalues are both real and unitary, and therefore either $+1$ or $-1$. Then, the operators $\mat{P}_\pm = \dfrac{1}{2}(\mat{I} \pm \mat{U})$ are orthogonal projectors into the eigenspaces of $\mat{U}$ to the eigenvalues $\pm 1$. Evidently, $\mat{U}$ %spectrally decomposes into 
can be expressed as $\mat{U}=\mat{P}_+ - \mat{P}_-$. Before measurement, the system is in the state

\begin{align}
\frac{1}{2} \Bigl[ \left( \ket{0}+\ket{1} \right) \otimes \mat{I} \ket{\psi_{in}}+ \left( \ket{0}-\ket{1} \right) \otimes \mat{U} \ket{\psi_{in}}\Bigr] \\
\quad = \ket{0}\otimes \mat{P}_+\ket{\psi_{in}} + \ket{1}\otimes \mat{P}_-\ket{\psi_{in}}.
\end{align}
A measurement of the first qubit in the standard basis, with the measurement operators 
$ \mat{P}_0 = \ket{0}\bra{0}\otimes \mat{I} $ and $ \mat{P}_1 = \ket{1} \bra{1} \otimes \mat{I} $, yields the probabilities
\begin{equation}
    p(0) = \braket{\psi_{in}|\mat{P}_+|\psi_{in}}
    \qquad \text{and} \qquad
    p(1) = \braket{\psi_{in}|\mat{P}_-|\psi_{in}},
\end{equation}
and leaves the system in the post-measurement state
\begin{equation}
     \ket{0} \otimes \dfrac{\mat{P}_+ \ket{\psi_{in}}}{\sqrt{\braket{\psi_{in}|\mat{P}_+|\psi_{in}}}}
\end{equation}
if the output $0$ is observed, respectively,
\begin{equation}
    \ket{1} \otimes \dfrac{\mat{P}_- \ket{\psi_{in}}}{\sqrt{\braket{\psi_{in}|\mat{P}_-|\psi_{in}}}}
\end{equation}
if the output $1$ is observed. 

Thus, the second register of the circuit is in post-measurement state of the observable $\mat{U}$ on a trial state $ \ket{\psi_{in}} $. Its expectation is

\begin{equation}
    \braket{\mat{U}} = 1\cdot p(0) + (-1)\cdot p(1) = \braket{\psi_{in}|(\mat{P}_+ - \mat{P}_-)|\psi_{in}}= \braket{\psi_{in}|\mat{U}|\psi_{in}},
\end{equation}
which shows that we can evaluate the expectation $\braket{\mat{U}}$ by measuring a single qubit. 
This is a crucial feature of our algorithm, as it make it possible to accurately approximate $\braket{\mat{U}}$ 
without having to sample $\bra{\psi_{in}}$ needed for the scalar product.
Also, it allows to use a moderate number of measurements that is independent of the number of qubits in $ \ket{\psi_{in}} $.
\vspace{1cm}
\end{figure*}
\end{appendices}





\end{document}